# Designs for a two-dimensional Si quantum dot array with spin qubit addressability

Masahiro Tadokoro[1,2], Takashi Nakajima[2], Takashi Kobayashi[3], Kenta Takeda[2], Akito Noiri[2], Kaito Tomari[1], Jun Yoneda[4], Seigo Tarucha[2,3], and Tetsuo Kodera[1,*]

1. Department of Electrical and Electronic Engineering, Tokyo Institute of Technology, Meguro-ku, Tokyo, 152-8552, Japan
2. Center for Emergent Matter Science, RIKEN, Wako-shi, Saitama, 351-0198, Japan
3. RIKEN Center for Quantum Computing, RIKEN, Wako-shi, Saitama, 351-0198, Japan
4. Tokyo Tech Academy for Super Smart Society, Tokyo Institute of Technology, Meguro-ku, Tokyo 152-8552, Japan
*Correspondence and requests for materials should be addressed to T.K. (email: kodera.t.ac@m.titech.ac.jp).

## Abstract

Electron spins in Si are an attractive platform for quantum computation, backed with their scalability and fast, high-fidelity quantum logic gates. Despite the importance of two-dimensional integration with efficient connectivity between qubits for medium- to large-scale quantum computation, however, a practical device design that guarantees qubit addressability is yet to be seen. Here, we propose a practical 3 × 3 quantum dot device design and a larger-scale design as a longer-term target. The design goal is to realize qubit connectivity to the four nearest neighbors while ensuring addressability. We show that a 3 × 3 quantum dot array can execute four-qubit Grover's algorithm more efficiently than the one-dimensional counterpart. To scale up the two-dimensional array beyond 3 × 3, we propose a novel structure with ferromagnetic gate electrodes. Our results showcase the possibility of medium-sized quantum processors in Si with fast quantum logic gates and long coherence times.

## Main text

### Introduction

Semiconductor spin qubits based on electron spins are an attractive candidate for a quantum processor[1,2]. As they can be manufactured using existing micro-fabrication techniques, it may be possible to integrate millions of qubits required for fault-tolerant quantum computation[3-5]. Electron spin qubits in Si in particular have attracted a good deal of attention in recent years because of their long coherence times and high-fidelity quantum logic gates and readout[6-27]. While one-dimensional qubit arrays have been employed in these pioneering experiments so far, some blueprints of two-dimensional integration have been presented for efficient connectivity between qubits[28-30]. However, proposals of feasible device designs that ensure the qubit addressability in a two-dimensional qubit array have been lacking despite its crucial role for scalable quantum computation[1].

In this study, we propose the designs of a two-dimensional Si quantum dot (QD) array with spin qubit addressability. First, we discuss a device design of a 3 × 3 QD array, which is the smallest two-dimensional square array with a QD connected to the four nearest neighbors, by largely relying on device fabrication processes established in academic laboratories[22,24]. We assess the effect of improved qubit connectivity against a 2 × 4 counterpart through the performance analysis of a four-qubit Grover's algorithm[31]. Next, as a longer-term approach, we propose a larger scale two-dimensional QD array with Co gates for fast spin manipulation and an additional Co magnet for qubit addressability. We introduce vias to enhance the QD density and connectivity, which is helpful for large quantum circuits. Our results present a potential pathway toward the development of quantum processors in Si comprising more than a thousand of spin qubits, with high-fidelity quantum logic gates and long coherence times.

## Results

### 3 × 3 QD array

Figure 1 illustrates our proposal of a 3 × 3 QD array. Figure 1(a) shows a model in which nine QDs are formed within a Si/SiGe heterostructure. We assume QD regions of 70-nm squares (red regions) separated by 50-nm barrier gates (deep blue regions) that are used to control exchange interactions between the adjacent QDs. The reservoirs (orange regions) connected to ohmic contacts (not shown) supply electrons to the QDs.

We consider two approaches to deplete electrons outside the QDs and reservoirs. The first one is etching followed by $SiO_2$ deposition and the second one is to use screening gates. Figure 1(b) shows a layer-by-layer schematic of overlapping Al gates[32] in the first approach, and Fig. 1(c) and (d) show vertical cross sections at the QD position and the barrier gate centers, respectively. In the first approach, a chemical mechanical polishing process may be used to obtain $SiO_2$ layer planarization[33,34], easing the deposition of four overlapping gate layers – two for barrier gates (the first- and third-layer gates) and the other two for plunger gates (the second- and fourth-layer gates). Table 1 summarizes the layer indexes, gate names, and colors in Figs. 1(b)-(d) and the respective gate sizes. In this approach, we can reduce the number of overlapping gate layers. On the other hand, we do not have to perform etching and $SiO_2$ deposition in the second approach. In the following, we consider the array architecture realized in the first approach.

We confirm the formation of electrostatic potentials for each QD in a Si quantum well (QW) in the room temperature simulation [Fig. 1(e)] carried out using COMSOL Multiphysics. It is seen from the figure that nine QDs are formed under every plunger gate, each of them is separated from the neighboring QDs and reservoirs by barrier gates.

### Qubit operation in a two-dimensional QD array

In order to utilize this QD array as a quantum processor, it needs to be capable of spin readout, initialization, and manipulation. While this device does not have dedicated change sensor QDs, spin readout can be performed by the gate-based sensing techniques to detect the Pauli spin blockade (PSB) between neighboring QDs[15,20]. The spin state can be initialized, for instance, by relaxation to the doubly occupied ground singlet state and rapid adiabatic passage[27]. For spin manipulation, we employ the electric-dipole spin resonance (EDSR) control based on micromagnets (MMs). The advantage of this scheme is that we can control the resonance frequency difference ($\Delta f_r$) between neighboring QDs with the MM design[35-37]. While spin manipulation can also be implemented with ESR striplines[7,8,14,25,27-30], the control of $\Delta f_r$ in this case relies on the difference of g-factors between QDs and it may be difficult to obtain controlled $\Delta f_r$ across a large QD array[7,8,14]. In contrast, MMs can potentially control $f_r$ and $\Delta f_r$ in a consistent manner across the qubit array by properly designing them, as we will show through simulations in the following.

Figure 2(a) shows the MM design for inducing the spatially inhomogeneous magnetic field ($B_{MM}$) in which the external magnetic field ($B_{ext}$) is applied in the direction of the arrow. The MM field $B_{MM}$ contains two essential components[12]: a transverse field ($b_{trans}$) perpendicular to $B_{ext}$ and a longitudinal field ($B_{long}$) parallel to $B_{ext}$. $b_{trans}$ enables spin rotations combined with a QD displacement ($\Delta x$) by inducing an effective oscillating magnetic field. On the other hand, $B_{long}$ provides the qubit addressability by shifting the Zeeman energy at each QD. Moreover, this QD-dependent energy shift is essential for implementing two-qubit gates such as the controlled-not gate[8,9], controlled-phase gate[8,10] and resonant SWAP gate[17,22]. For high fidelity single-qubit gates, $f_{Rabi}$ faster than 1 MHz is desirable[9,12].

Table 2(a) and Figure 2(b) show the simulated transverse field slope ($b_{trans}$) at the QD positions. $b_{trans}$ at each QD position is calculated to be 0.56 – 1.2 mT/nm, comparable to the values in the previously demonstrated linear QD arrays[12]. We estimate $f_{Rabi}$ to be 6.8 – 14 MHz assuming a conversion factor from $b_{trans}$ to $f_{Rabi}$ of 12 MHz·nm/mT, which is taken from previous experiment[9]. We assume that the MW drive is applied to the barrier gate next to the QD (along $B_{ext}$) to maximize $f_{Rabi}$. In order to assess the qubit addressability, we calculate $B_{long}$ (Fig. 2(c) and Table 2(b)). The minimum $B_{long}$ difference ($\Delta B_{long}$) over the set of nine QDs is 6 mT (between $Q_{23}$ and $Q_{33}$) and it corresponds to $\Delta f_r$ of 160 MHz. This is much larger than our calculated $f_{Rabi}$ and therefore large enough to prevent the crosstalk of single-qubit gates. These results show this MM design can induce field gradients necessary for the qubit addressability even in a 3 × 3 QD array.

### Efficient quantum circuit execution in a two-dimensional QD array

One of the potential advantages of two-dimensional qubit arrays is the reduced circuit depth of quantum algorithm implementations thanks to better qubit connectivity[38]. Here, we discuss the efficiency of the four-qubit Grover's search algorithm[31] implemented in our two-dimensional QD array, see Fig. 3(a). This algorithm allows one to search for $|q_3 q_2 q_1 q_0\rangle = |1101\rangle$ using the superposition of four qubits with the same probability (we define $|0\rangle$ by the spin down state). The four-qubit Toffoli gates play a central role in implementing this circuit and can be synthesized from 15 two-qubit gates including 2 SWAP gates for compensating the lack of direct qubit couplings [Fig. 3(b)][39,45] (see Methods for the details). The spin state of qubit qX is read out by PSB with a measurement ancilla qubit $M_{qX}$. For comparison, we consider a linear array of four QDs which are lined up together with adjacent QDs for spin readout as shown in Fig. 3(c). In this qubit layout, 18 two-qubit gates including 5 SWAP gates are necessary to implement a four-qubit Toffoli gate. This shows that the two-dimensional QD array enables efficient quantum circuit execution due to improved qubit connectivity.

### A larger scale two-dimensional array

The scalability of the EDSR control based on MMs has often been questioned because of its difficulty in inducing strong field gradients over large areas[30]. We have nevertheless shown that it is possible to induce field gradients in a small two-dimensional QD array. In the following, we propose a longer-term design approach, as outlined in Fig. 4(a). This structure may be amenable to a larger qubit array with fast, MM-mediated electrical spin control at the expense of introducing vias to enhance QD density and connectivity as compared to the 3 × 3 QD array. Under this structure, $b_{\text{trans}}$ and $B_{\text{long}}$ will be produced separately: $b_{\text{trans}}$ is induced by the plunger and barrier gate electrodes made of Co instead of Al, whereas $B_{\text{long}}$ is induced by a large Co magnet located outside of the QD array.

Under this novel magnet structure, it would be possible to integrate about 40 × 40 = 1600 QDs within a 5 × 5-$\mu m^2$ area because each QD occupies only 120 × 120-$nm^2$. Assuming that PSB is used for spin state initialization as discussed above, we can consider an example of implementing qubits and ancillas in this QD array as shown in Fig. 4(b). In this example, the 3 × 3 QD array consisting of 4 data qubits (red) and 4 ancillas functions as a unit cell. QDs with a cross are not used, so they can be empty, in which case coherent inter-site qubit transport[25] may be used to maintain high qubit connectivity. Alternatively, in the case of single occupancy, two-qubit exchange gates[17] would be needed for connection. As can be seen from this example, such a device would be able to host more than a thousand of qubits, a number of qubits much larger than those in any other existing quantum processors[40-44], in a way compatible with high qubit connectivity and the spin operation scheme discussed above.

Figures 4(c) and (d) show simulated values of $B_{\text{long}}$ slope in the $x$ and $y$ directions induced by a large Co magnet located outside the QD array (yellow square). In the QD array, the $\Delta f_r$ values between the nearest-neighbor QDs are larger than 100 MHz (corresponding to >0.03 mT/nm), which are sufficient for unconditional single-qubit operations and two-qubit gate manipulations. Figure 4(e) shows simulated values of $b_{trans}$ induced by the Co gates. Here, we use a 5 × 5 QD array for simulation simplicity and assume the gate sizes of Co plunger and barrier gates are 60 nm × 60 nm and 60 nm × 40 nm, respectively. The estimated average value of $f_{Rabi}$ for the 25 QDs is 39 MHz (assuming the same conversion factor used in the previous discussion), which is about three times faster than that in the 3 × 3 array discussed above owing to the close proximity of the magnet and QDs. By applying the proposed magnet fabrication approach, quantum processors with large numbers of qubits may be achieved with high-speed qubit manipulation, leveraging mature fabrication and integration techniques in Si. We anticipate that even larger arrays will be possible by further improving the magnet design.

## Discussion

In this study, we discuss the designs of a two-dimensional Si QD array for multi-qubit quantum processing with qubit addressability. We reveal the possibility to provide qubit addressability in the two-dimensional array of 3 × 3 QDs, which can execute quantum circuits more efficiently compared with the one-dimensional array. We note that this method is essentially based on experimentally realized MM and QD structures. We have furthermore presented a possibility of hosting hundreds of qubits with $f_{Rabi}$ increased by roughly three times by employing a novel magnet-incorporated QD structure. We believe both methods will allow to scale up QD array sizes, to a level that was previously thought to be difficult with this scheme. Our results have shown how to guarantee qubit addressability when electron spin qubits in Si are integrated to a larger scale. Quantum processors with large numbers of qubits, high-fidelity quantum logic gates, and long coherence times might accelerate research on quantum algorithms and architectures, enhancing the potential of quantum computing.

## Methods

### The method for mapping quantum circuits to QD arrays

We employed Qiskit's LookaheasSwap routing method[45] for mapping quantum circuits via insertion of SWAP gates to construct quantum circuits for four-qubit Grover's search algorithm. This method repeats the process of selecting the SWAP gate that maximizes the number of subsequent executable two-qubit gates when there is a two-qubit gate that cannot be executed in the current physical qubit layout. While Qiskit provides other methods to map circuits by inserting SWAP gates, we chose this method because it produces the circuit with smaller numbers of SWAP gates for both two-dimensional and one-dimensional QD arrays.

## References
1. DiVincenzo, D. P. The physical implementation of quantum computation. *Fortschritte der Physik: Progress of Physics* **48,** 771 (2000).
2. Ladd, T. D. *et al.* Quantum computers. *Nature* **464,** 45 (2010).
3. Ito, T. *et al.* Four single-spin Rabi oscillations in a quadruple quantum dot. *Applied Physics Letters* **113,** 093102 (2018).
4. Mills, A. R. *et al.* Shuttling a single charge across a one-dimensional array of silicon quantum dots. *Nature Communication* **10,** 1063 (2019).
5. Mortemousque, P.A. *et al.* Coherent control of individual electron spins in a two-dimensional quantum dot array. *Nature Nanotechnology* (2020).
6. Loss, D., DiVincenzo, D. P. Quantum computation with quantum dots. *Physical Review A* **57,** 120 (1998).
7. Veldhorst, M. *et al.* An addressable quantum dot qubit with fault-tolerant control-fidelity. *Nature nanotechnology* **9,** 981 (2014).
8. Veldhorst, M. *et al.* A two-qubit logic gate in silicon. *Nature* **526,** 410 (2015).
9. Takeda, K. *et al.* A fault-tolerant addressable spin qubit in a natural silicon quantum dot. *Science advances* **2,** e1600694 (2016).
10. Watson, T. F. *et al.* A programmable two-qubit quantum processor in silicon. *Nature* **555,** 633 (2018).
11. Zajac, D. M. *et al.* Resonantly driven CNOT gate for electron spins. *Science* **359,** 439 (2018).
12. Yoneda, J. *et al.* A quantum-dot spin qubit with coherence limited by charge noise and fidelity higher than 99.9%. *Nature nanotechnology* **13,** 102 (2018).
13. Takeda, K. *et al.* Optimized electrical control of a Si/SiGe spin qubit in the presence of an induced frequency shift. *npj Quantum Information* **4,** 1 (2018).
14. Huang, W. *et al.* Fidelity benchmarks for two-qubit gates in silicon. *Nature* **569,** 532 (2019).
15. Zheng, G. *et al.* Rapid gate-based spin read-out in silicon using an on-chip resonator. *Nature nanotechnology* **14,** 742 (2019).
16. Volk, C. *et al.* Fast charge sensing of Si/SiGe quantum dots via a high-frequency accumulation gate. *Nano letters* **19,** 5628 (2019).

17. Sigillito, A. J. *et al.* Coherent transfer of quantum information in a silicon double quantum dot using resonant SWAP gates. *npj Quantum Information* **5,** 1 (2019).
18. Noiri, A. *et al.* Radio-Frequency-Detected Fast Charge Sensing in Undoped Silicon Quantum Dots. *Nano Letters* **20,** 947 (2020).
19. Leon, R. C. C. *et al.* Coherent spin control of s-, p-, d- and f-electrons in a silicon quantum dot. *Nature communications* **11,** 1 (2020).
20. Schaal, S. *et al.* Fast gate-based readout of silicon quantum dots using Josephson parametric amplification. *Physical Review Letters* **124,** 067701 (2020).
21. Yoneda, J. *et al.* Quantum non-demolition readout of an electron spin in silicon. *Nature communications* **11,** 1 (2020).
22. Takeda, K. *et al.* Resonantly driven singlet-triplet spin qubit in silicon. *Physical Review Letters* **124,** 117701 (2020).
23. Yang, C. H. *et al.* Operation of a silicon quantum processor unit cell above one kelvin. *Nature* **580,** 350 (2020).
24. Takeda, K. *et al.* Quantum tomography of an entangled three-spin state in silicon. *arXiv preprint arXiv:2010.10316* (2020).
25. Yoneda, J. *et al.* Coherent spin qubit transport in silicon. *arXiv preprint arXiv:2008.04020* (2020).
26. Bohuslavskyi, H. *et al.* Reflectometry of charge transitions in a silicon quadruple dot. *arXiv preprint arXiv:2012.04791* (2020).
27. Seedhouse, A. *et al.* Pauli Blockade in Silicon Quantum Dots with Spin-Orbit Control. *PRX Quantum* **2,** 010303 (2021).
28. Vandersypen, L. M. K. *et al.* Interfacing spin qubits in quantum dots and donors—hot, dense, and coherent. *npj Quantum Information* **3,** 1 (2017).
29. Veldhorst, M. *et al.* Silicon CMOS architecture for a spin-based quantum computer. *Nature communications* **8,** 1 (2017).
30. Li, R. *et al.* A crossbar network for silicon quantum dot qubits. *Science advances* **4,** eaar3960 (2018).
31. Grover, L. K. Quantum mechanics helps in searching for a needle in a haystack. *Physical Review Letters* **79,** 325 (1997).
32. Angus, S. J. *et al.* Gate-defined quantum dots in intrinsic silicon. *Nano Letters* **7,** 2051 (2007).
33. Kaufman, F. B. *et al.* Chemical-mechanical polishing for fabricating patterned W metal features as chip interconnects, *Journal of the Electrochemical Society* **138,** 3460 (1991).
34. Song, L. *et al.* Influences of silicon-rich shallow trench isolation on total ionizing dose hardening and gate oxide integrity in a 130 nm partially depleted SOI CMOS technology. *Microelectronics Reliability* **74,** 1 (2017).
35. Tokura, Y. *et al.* Coherent single electron spin control in a slanting Zeeman field. *Physical Review Letters* **96,** 047202 (2006).
36. Pioro-Ladrière, M. *et al.* Electrically driven single-electron spin resonance in a slanting Zeeman field. *Nature Physics* **4,** 776 (2008).
37. Yoneda, J. *et al.* Robust micromagnet design for fast electrical manipulations of single spins in quantum dots. *Applied Physics Express* **8,** 084401 (2015).
38. Linke, N. M., *et al.* Experimental comparison of two quantum computing architectures. *Proceedings of the National Academy of Sciences* **114,** 3305 (2017).
39. Barenco, A. *et al.* Elementary gates for quantum computation. *Physical review A* **52,** 3457 (1995).
40. Friis, N. *et al.* Observation of Entangled States of a Fully Controlled 20-Qubit System. *Physical Review X* **8,** 021012 (2018).
41. Bradley, C.E. *et al.* A Ten-Qubit Solid-State Spin Register with Quantum Memory up to One Minute. *Physical Review X* **9,** 031045 (2019).

**42.** Arute, F. *et al.* Quantum supremacy using a programmable superconducting processor. *Nature* **574,** 505 (2019).

**43.** Zhong, H.-S. *et al.* Quantum computational advantage using photons. *Science* **370,** 1460 (2020).

**44.** Ebadi, S. *et al.* Quantum Phases of Matter on a 256-Atom Programmable Quantum Simulator. *arXiv preprint arXiv:2012.12281* (2020).

**45.** Jandura's routing method (LookaheadSwap documentation). https://qiskit.org/documentation/stubs/qiskit.transpiler.passes.LookaheadSwap.html#qiskit.transpiler.passes.LookaheadSwap (accessed June 10, 2021).

**Acknowledgements**
The authors thank Dr. B. S. Choi for valuable discussions regarding quantum circuit analysis. This work was supported financially by Core Research for Evolutional Science and Technology (CREST), Japan Science and Technology Agency (JST) (JPMJCR1675), JST Moonshot R&D grant no. JPMJMS2065, JST PRESTO grant no. JPMJPR2017, MEXT Quantum Leap Flagship Program (MEXT Q-LEAP) grant no. JPMXS0118069228, and JSPS KAKENHI grant nos. 18H01819, 18K18996, 20H00237 and 21K14485.

**Author Contributions**
M.T., T.N., T.K., K.T., and A.N. contributed to the device design. M.T. and K.T. performed simulations. M.T. wrote the main manuscript text with input from all authors. J.Y. critically reviewed the manuscript and assisted with revisions. S.T. and T.K. supervised the project.

**Additional Information**
**Competing Interests:** The authors declare no competing interests.

Figures and tables

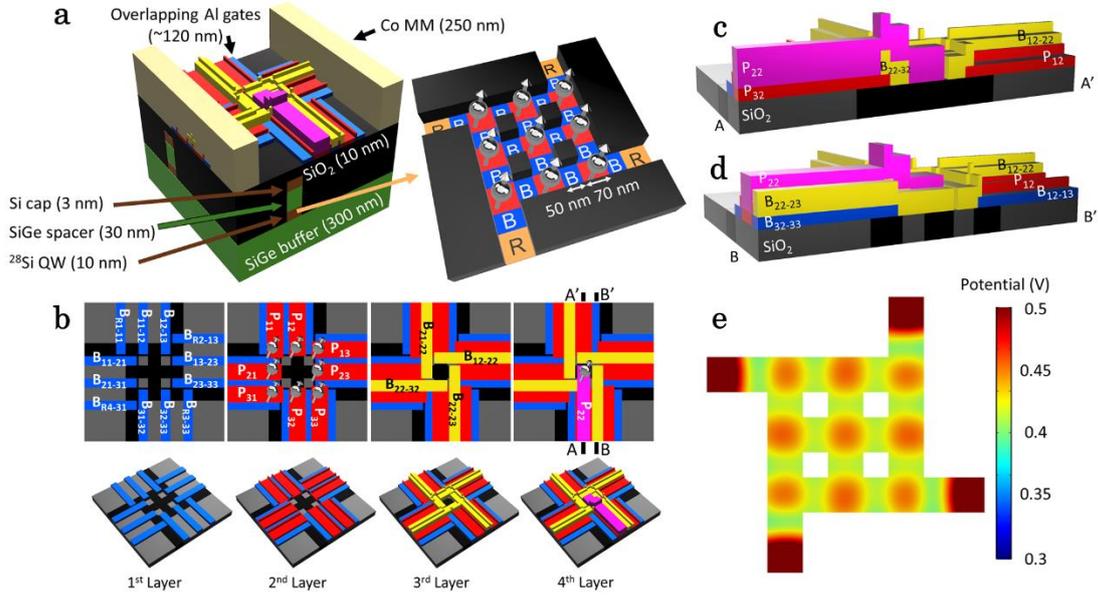

**Figure 1. Schematics of a device design with a 3 × 3 QD array** (a) Device layer structure and 3 × 3 QD array schematic in Si QW. Silver spheres with arrows represent spins in QDs. Labels B and R in the array represent barrier gates and reservoirs, respectively. (b) Layer stack of overlapping Al gates. The electron spin symbol is displayed when the corresponding control gate electrode is fabricated in the layer. This device can be fabricated with four layers of overlapping gates. An overlay accuracy of 10 nm is assumed. (c),(d) Cross sections of the QDs (A-A') and barrier gates (B-B') in (b). The gate thickness increases in the upper layers to ensure gate continuity. In (d), the black and gray areas are both $SiO_2$, with the Si layer underneath etched in the latter. (e) Electrostatic potential in the Si QW at applied voltages of 0.6 V on plunger gates and 0.4 V (0.3 V) on the barrier gates between QDs (between a QD and a reservoir).

**Table 1. Overlapping-layer gate characteristics.**

| Layer index | Gate name | Gate color | Gate width (nm) | Gate height (nm) |
|---|---|---|---|---|
| 1 | $B_{11-12}$, $B_{12-13}$… $B_{r1-11}$, $B_{r2-13}$… | Blue | 50 | 15 |
| 2 | $P_{11}$, $P_{12}$… | Red | 90 | 25 |
| 3 | $B_{12-22}$, $B_{21-22}$… | Yellow | 60 | 40 |
| 4 | $P_{22}$ | Magenta | 70 | 60 |

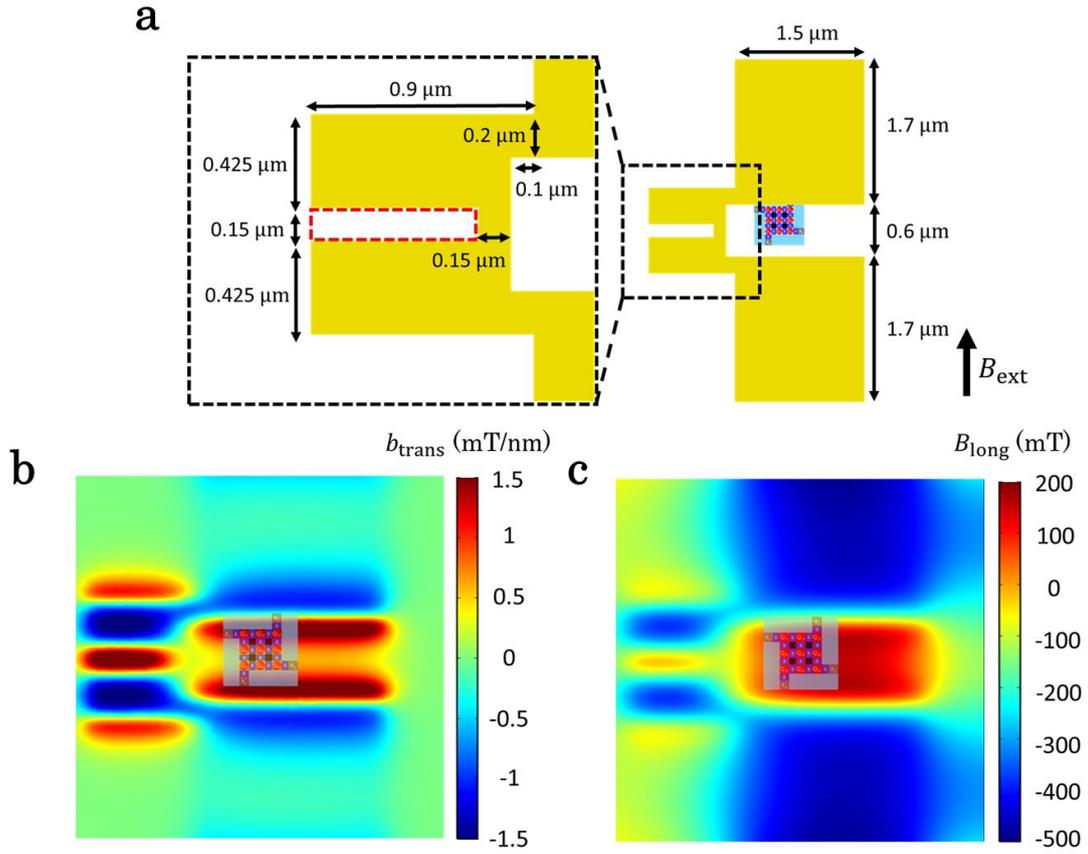

**Figure 2. The schematics of MM design and simulation results.** (a) MM design. The right panel illustrates the layout of the MM (yellow) along with the 3 x 3 QD array (center), designed so that the values of $\Delta f_r$ are robust against misalignment between the QD array and the MM. The magnet is 250 nm thick and 143 nm apart from the bottom of the magnet to the QD. Left panel shows an enlarged view of the region surrounded by the black dashed rectangle in the right. The magnet has a groove in the middle with its shape shown by the red dashed rectangle to make $\Delta f_r$ in each QD row moderate. $B_{\text{ext}}$ indicates the direction of the external magnetic field. (b),(c) Simulation results for $b_{\text{trans}}$ and $B_{\text{long}}$ in the Si QW. Magnetization of the MM is 1,400 kA/m in the direction parallel to $B_{\text{ext}}$.

**Table 2(a).** $b_{trans}$ and $f_{Rabi}$ at each QD position.

| $b_{trans}$ (mT/nm) $f_{Rabi}$ (MHz) | Column 1 | Column 2 | Column 3 |
|---|---|---|---|
| Row 1 | 1.1<br>13 | 1.1<br>14 | 1.2<br>14 |
| Row 2 | 0.56<br>6.8 | 0.60<br>7.2 | 0.62<br>7.4 |
| Row 3 | 0.73<br>8.7 | 0.76<br>9.2 | 0.77<br>9.3 |

**Table 2(b).** $B_{long}$ and $\Delta f_r$ at each QD position; $\Delta f_r$ is calculated with respect to the center QD.

| $B_{long}$ (mT) $\Delta f_r$ (MHz) | Column 1 | Column 2 | Column 3 |
|---|---|---|---|
| Row 1 | 110<br>-370 | 140<br>390 | 150<br>880 |
| Row 2 | 96<br>-770 | 120<br>0 | 140<br>510 |
| Row 3 | 100<br>-600 | 130<br>170 | 150<br>670 |

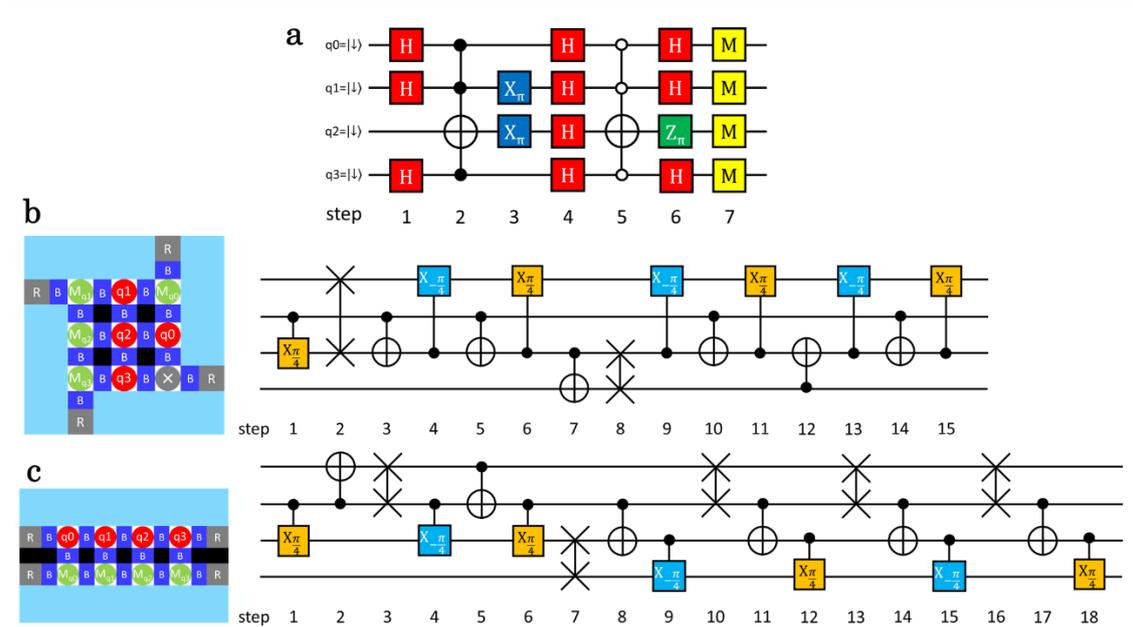

**Fig. 3. Quantum circuits for four-qubit Grover's search algorithm.** (a) Quantum circuit for implementing a four-qubit Grover's search algorithm. H, X, Z, and M indicate Hadamard gate, Pauli-X gate, Pauli-Z gate, and spin state measurement, respectively. This circuit is optimized to reduce the number of quantum gates[39]. Four qubits are prepared in an equally weighted superposition state and then the probability of $|q_3 q_2 q_1 q_0\rangle = |1101\rangle$ is amplified once. (b),(c) Implementation examples of a four-qubit Toffoli gate in two-dimensional (b) and one-dimensional (c) QD arrays using SWAP and controlled rotation gates (see Methods). Four red QDs are used for computation and four green QDs host ancillas for spin state initialization and measurement using PSB. Bottom-right QD in (b) is not used in this example.

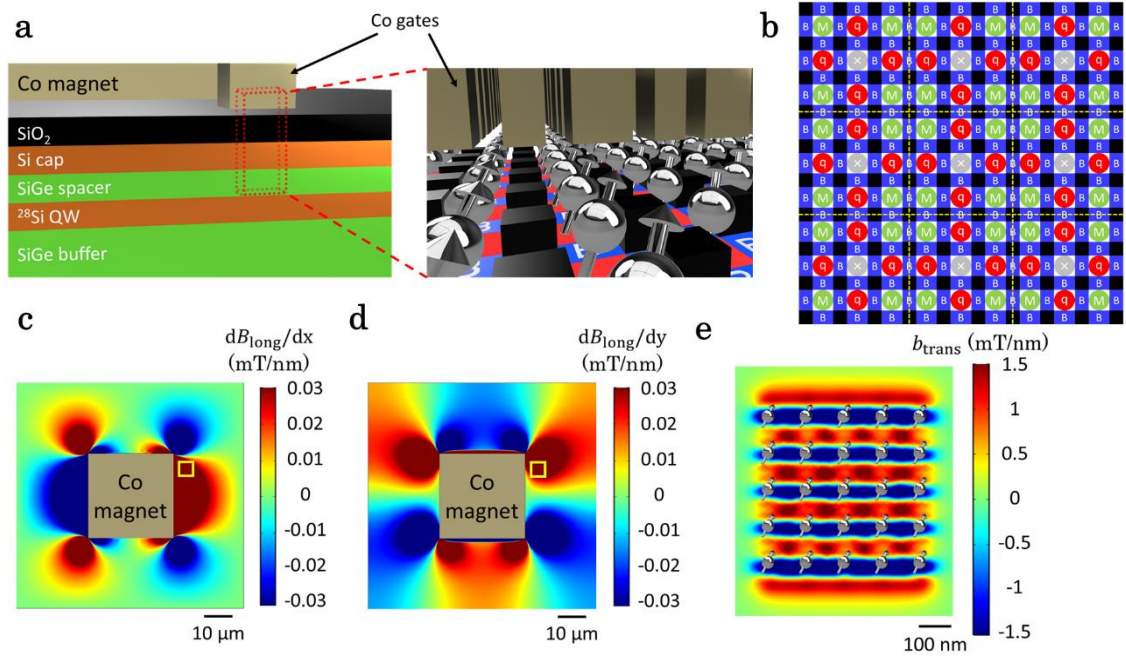

**Fig. 4. Quantum processor with larger numbers of QDs with novel magnet structure.**
(a) Model of a quantum processor utilizing a novel magnet structure with a large Co magnet and Co gates. The right figure shows a zoom-in of the red dotted area in the left figure. The whole 5 × 5-μm² array contains approximately 1600 QDs. (b) Implementation example of qubits and ancillas for spin state initialization and measurement in a larger two-dimensional array. It only shows a 9 × 9 QD array, but we can straightforwardly scale it up to a larger one. (c),(d) Simulated values of $B_{\text{long}}$ slope in the *x* and *y* directions induced by a large Co magnet (30 × 30 × 5 μm³) located outside the QD array. The yellow square indicates the position of a 5 × 5-μm² QD array, in which $\Delta B_{\text{long}}$ is sufficiently large. By tuning the magnet aspect ratio, it may be possible to control the distribution of the magnetic field so that a larger QD array may be accommodated. (e) Simulated values of $b_{\text{trans}}$ induced by Co gates. For simulation simplicity, a 5 × 5 QD array is shown, with the positions of spin qubits illustrated by symbols.